\documentclass[prx,twocolumn,superscriptaddress,amsmath,amssymb,floatfix]{revtex4}
\usepackage{graphicx,color}
\usepackage{dcolumn}
\usepackage{bm}
\usepackage{setspace}

\usepackage{color}

\begin{document}

\title{Tunable coupling to a mechanical oscillator circuit using a coherent feedback network}
\author{Joseph Kerckhoff} \email{jkerc@jila.colorado.edu}
\address{JILA, National Institute of Standards and Technology and the University of Colorado, Boulder, Colorado 80309, USA}
\author{Reed W. Andrews}
\address{JILA, National Institute of Standards and Technology and the University of Colorado, Boulder, Colorado 80309, USA}
\author{H. S. Ku}
\address{JILA, National Institute of Standards and Technology and the University of Colorado, Boulder, Colorado 80309, USA}
\author{William F.  Kindel}
\address{JILA, National Institute of Standards and Technology and the University of Colorado, Boulder, Colorado 80309, USA}
\author{Katarina Cicak}
\address{National Institute of Standards and Technology, Boulder, CO 80305, USA}
\author{Raymond W. Simmonds}
\address{National Institute of Standards and Technology, Boulder, CO 80305, USA}
\author{K. W. Lehnert}
\address{JILA, National Institute of Standards and Technology and the University of Colorado, Boulder, Colorado 80309, USA}

\date{\today}
\pacs{43.35.Gk,84.30.Bv,74.81.Fa,42.50.Ct}

\begin{abstract}
We demonstrate a fully cryogenic microwave feedback network composed of modular superconducting devices connected by transmission lines and designed to control a mechanical oscillator coupled to one of the devices.  The network features an electromechanical device and a tunable controller that coherently receives, processes and feeds back continuous microwave signals that modify the dynamics and readout of the mechanical state.  While previous electromechanical systems represent some compromise between efficient control and efficient readout of the mechanical state, as set by the electromagnetic decay rate, the tunable controller produces a closed-loop network that can be dynamically and continuously tuned between both extremes much faster than the mechanical response time.  We demonstrate that the microwave decay rate may be modulated by at least a factor of 10 at a rate greater than $10^4$ times the mechanical response rate.  The system is easy to build and suggests that some useful functions may arise most naturally at the network-level of modular, quantum electromagnetic devices.
\end{abstract}

\maketitle

\section{\label{sec:Intro} Introduction}
As researchers improve their ability to engineer quantum electromagnetic (EM) devices, we should start to look beyond individual components and find experimental and theoretical techniques that exploit coherent networks of quantum devices \cite{G&J}.  For instance, it is important to consider whether useful functions are best achieved at the {\it network-level} of interacting, modular, and increasingly generic devices.  This approach is routine in classical electronics.  For example, an unreliable operational amplifier (``op-amp'') may be combined with a reliable impedance network in a feedback configuration to realize a reliable amplifier.  This strategy is usually more efficient than developing high quality amplifier chips for each new use.  If quantum engineering is to consider this approach, it would greatly benefit from some systemization of device and interconnection laws, in very rough analogy to Kirchoff's laws' systemization of practical electronics.    Because classical network theory has so many established, practical techniques, we might expect that hybrid approaches towards this end may handle network complexity better than approaches entirely native to physics.

Feedback engineering addresses two types of problems: feedback can enhance a system's robustness and/or tune the dynamics of an otherwise untunable system \cite{AM,WM}.  As a general rule, modern quantum EM devices feature increasingly monolithic and insular designs, as robustness is generally prized over tunability in single-device physics.  In the field of electromechanics (also known as ``optomechanics'') for example, devices in which mechanical simple harmonic oscillators are coupled to itinerant EM fields have become increasingly powerful, e.g. \cite{Kipp08,OCon10,Teuf11,Chan11,Palo13,Purd13,Safa12}.  Mechanical oscillators are more isolated from thermal baths, interactions with distinct EM modes are better controlled and quantum effects are starting to be observed \cite{OCon10,Teuf11,Chan11,Palo13,Purd13,Safa12}.  But as a consequence, these devices are often module-like and unadjustable: important characteristics like EM center frequency and linewidth are often fixed by construction, and often a single EM input/output (I/O) port admits access to all internal mechanisms \cite{Teuf11,Chan11,Safa12,Palo13}.  In principle, tunability may be won back without compromising device integrity by employing a tunable, coherent controller that exchanges continuous coherent signals with this port.  Just as a reliable, negative feedback network can ameliorate the unreliable gain of an op-amp, perhaps combining a tunable coherent feedback network with an insulated electromechanical circuit may bring out the best of both, and at low cost.

Here we demonstrate a coherent feedback network of EM devices, a superconducting electromechanical device \cite{Teuf08,LaHa09,Hertz10,OCon10,Teuf11,Mass12,Rega11,Andr13,Teuf09} and a superconducting microwave controller \cite{Kerc12,Cast08}, that provides us with a type of dynamic flexibility previously unavailable in electromechanics.   Namely, while previous electromechanical systems make some compromise between better mechanical control and better mechanical measurement capabilities, we demonstrate that this network may be dynamically modulated between both extremes.  While our network is too complex for traditional electromechanical models \cite{Wils07,Marq07,Kipp08}, it requires only three components, all of which are accessible to superconducting circuit labs (and feature regularly in our own work \cite{Teuf09,Teuf11,Palo13}), and is efficiently and intuitively modeled as a linear feedback system \cite{AM,G&J,Goug10}.  In other words, this network is easy to build (given familiar technologies, specifically, devices ``generic'' in our lab) and its mechanisms easy to intuit.  But it is difficult to model quantitatively, meriting methods that naturally handle complex networks.  We do not characterize performance in the quantum regime, but these devices feature regularly in such work and these capabilities are novel regardless.  Moreover, because the network is linear and driven by Gaussian fields, the network characterization done in the classical regime should apply in the quantum one \cite{G&J,Hinf,LQG,Goug10,Hame12,Mabu08,Iida11}.  Thus, we use the term ``coherent'' in the sense that an interferometer or resonator is coherent, the primary prerequisite for quantum dynamics.

 Because a systematic network theory perspective is relatively novel in electromechanics \cite{Hame12,Bott12,Tsan10}, this article features a pedagogical modeling section that outlines the modeling method and its advantages (rigorous derivations are well-covered in the literature, e.g. \cite{G&J,Hinf,LQG,Goug10,Hame12,Wise94,Kerc10}).  Also, because it employs theoretical approaches developed by an electrical engineering community, familiar concepts in electromechanics are described in more detail than usual.  The outline of the paper is as follows, in section~\ref{sec:Network} we introduce the network construction and its physical intuition.  Section~\ref{sec:Results} presents our measurements of this network, its essential operation and agreement with the network model.  Section~\ref{sec:Model} describes the procedure for constructing the network model and section~\ref{sec:Con} concludes.  

\section{\label{sec:Network} The network and physical intuition}

Many superconducting microwave realizations of electromechanical systems consist of a high quality microwave LC resonator (hereon ``resonator''), in which the capacitance is modulated by a spring-loaded, mechanical simple harmonic oscillator (hereon ``MO'') \cite{Teuf08,Teuf09,Teuf11,Hertz10,Mass12,Rega11,Andr13}.  Motion in this MO adjusts the capacitance of the resonator, and thus acts on microwave power in the resonator.  And the voltage potential across the capacitor creates a Coulomb force that acts on the MO, and thus microwave power in the resonator acts on the position of the MO.  Via an inductive transformer (or some other ``input coupler''), this high-quality resonator couples to a transmission line, which serves as the I/O port through which the electromechanical device is probed, driven and controlled.  These microwave systems are accurately described at a quantum level with a model that is formally equivalent to the models used to describe electromechanical systems of all scales and realizations \cite{Brow07,Teuf08,Wils07,Marq07}.  

The choice of an input coupler in an electromechanical system typically represents a compromise.  For example, microwave signals inside a resonator leave the resonator more slowly if it couples to a transmission line weakly.  As a consequence, microwave signals in resonators with weak couplers both affect and are affected by the MO more strongly: the weak coupling causes electromechanical effects to integrate for a longer time.  But too much integration can also be undesirable.  For example, weak couplers frustrate high-bandwidth detection of mechanical motion by low-pass filtering mechanical information on EM signals leaving the resonator and encouraging loss of the same through parasitic channels.  The usual distinction between these two regimes is whether or not the resonator linewidth $\kappa_t$ is narrower than the MO's center frequency $\Omega$.  More strongly (weakly) coupled systems with $\kappa_t>\Omega$ ($\kappa_t<\Omega$) are known as ``unresolved sideband'' (``resolved sideband'') systems.  Experimentalists typically use resolved sideband systems to control and prepare specific states in MOs \cite{Teuf08,Teuf11,Chan11,Palo13}.  Unresolved sideband systems are most useful when high-bandwidth readout of the MO's motion is the priority \cite{Klec06,Gava12}.  Sometimes, intermediate regimes are ideal ($\kappa_t\approx\Omega$) \cite{Purd13,Teuf09}.  In all electromechanical systems to our knowledge, though, the resonator's coupling strength is fixed at the time of construction.      

How might one construct a more flexible system that could be tuned dynamically between the resolved and unresolved sideband regimes?  Our solution was inspired by considering what usually goes ``outside'' EMC devices.  Many superconducting microwave electromechanical experiments conducted by us \cite{Teuf08,Teuf11,Teuf09,Palo13} and others \cite{Brow07,LaHa09,Hertz10,Mass12} are partitioned into an ``upstream'' electromechanical circuit (EMC) and a ``downstream'' cryogenic low- or near quantum limited-noise temperature amplifier, for high-fidelity readout of microwave probes.  Our group typically uses near quantum-limited Josephson parametric amplifiers (JPAs) for readout \cite{Cast08}.  JPAs are composed of a 20 dB directional coupler followed by a single-port ``tunable Kerr circuit'' (TKC) \cite{Cast08,Kerc12}, a nonlinear microwave resonator whose center frequency is drive power-dependent and tunable with an applied magnetic flux.  The essential novelty in this work is that rather than measure the signal that comes out of the JPA, a portion of the signal emitted by the EMC and passed to the JPA is fed back into the EMC coherently.  This feedback gives microwave signals more opportunities to interact with the OM, which is the essential quality of a weakly-coupled/resolved-sideband system.  Moreover, because the JPA is dynamically tunable, we can choose whether or not this feedback occurs, and access both coupling regimes dynamically.  And because the critical components, the EMC and JPA, are ``generic'' in our lab, we can pursue this scheme using existing devices.  

\begin{figure}[b!]
\includegraphics[width=.4\textwidth]{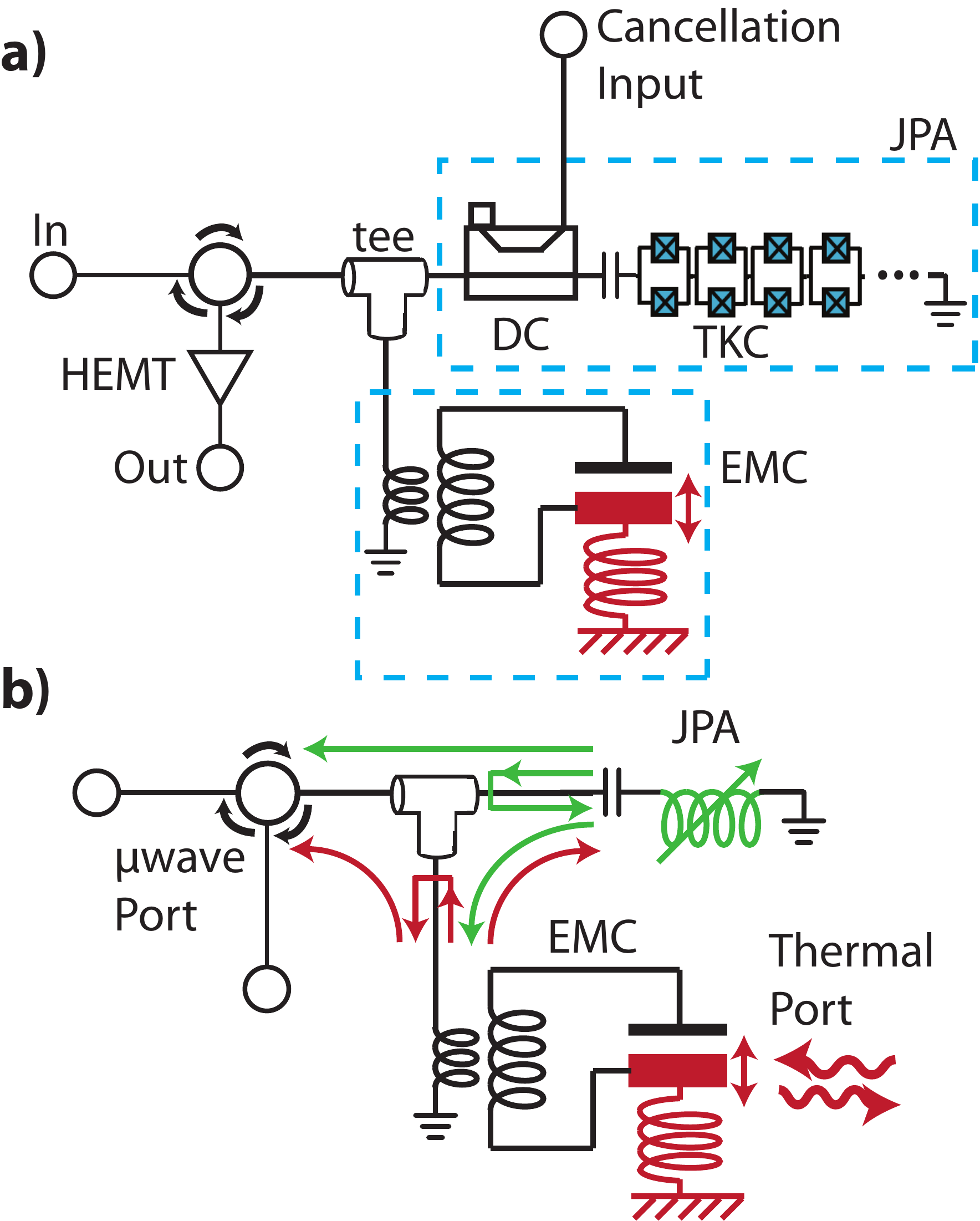}
\caption{\label{fig1} a) An experimental schematic of the coherent feedback network.  A microwave tee connects an electromechanical circuit (EMC) to a Josephson parametric amplifier (JPA) --- here broken down into its directional coupler (DC) and tunable Kerr circuit (TKC) subcomponents --- and provides an overall input-output port.  The In port is typically driven by a strong electromechanical coupling tone, but this tone is typically cancelled at the DC, preventing its strong carrier from driving the TKC nonlinear.  Microwave signals are detected at the Out port.  b) Conceptual schematic of coherent interconnections.  Ideally, the JPA acts as an over-coupled and wide-band linear resonator with a tunable center frequency.}
\vspace{-0.1in}
\end{figure}

Housed in a dilution refrigerator, our network is the interconnection of a single-port EMC, a JPA, and a commercial microwave tee \footnote{Specifically, the tee was a commonplace, commercial SMA coaxial adapter tee (all male), equivalent to those available from Pasternack (\# PE9379).} (Fig.~\ref{fig1}a).  The EMC and JPA are connected to one tee port each, while the remaining tee port serves as the overall I/O network port.  The network output is amplified by a cryogenic HEMT for analysis.  While the EMC and TKC were originally designed for different experiments and were mounted in separate sample boxes, and interconnections were made using $\sim$cm-length coplanar waveguides and low-loss, semi-rigid coaxial cables, an analogous network could be fabricated on a single Si substrate.  Although the specific electromechanical circuit (EMC) used here is somewhat incidental, it is a new design of this type \cite{Andr13}, built from a 4.672 GHz center frequency and over-coupled $\kappa_t/2\pi=2.8$ MHz lumped-element resonator, employing a mechanical membrane oscillator with an effective mass of $\sim$10 ng, center frequency of $\Omega=2\pi\times$713.6 kHz and intrinsic linewidth of $\Gamma_0=2\pi\times$0.81 Hz, and in which each photon exerts a force on the MO (in coherent amplitude units) of $g_0=h\times2.3$ Hz.  In this report, we do not pump our JPA; effectively, it is operated as a gain-1 amplifier, or more precisely, as a linear resonator whose center frequency can be tuned by applying either magnetic flux or a moderate amount of microwave drive power.  We do, however, typically employ a cancellation tone that prevents the carrier of any strong, electromechanical coupling tone driving the network from reaching the TKC and driving it nonlinear.  

The network's operation is depicted conceptually in Fig.~\ref{fig1}b.  Microwave signals leaving the EMC and carrying information about the motion of the MO are split three ways at the tee: most of the amplitude is split evenly between the network output and the JPA input, and a small amount is reflected.  Modeling the JPA as a linear resonator with a tunable center frequency, the JPA reflects its signal portion with a tunable phase shift.  Similarly split by the tee, a portion of the JPA-reflected signal interferes with the EMC-to-output signal, enhancing or diminishing the rate of mechanical information leaving the network.  This same interference also enhances or diminishes the effective microwave linewidth of the network, as seen by the microwave input port.  Finally, the portion of the JPA-reflected signal fed back to the EMC can exert a force on the MO, amplifying or counteracting its motion.  All of these effects are controlled by a single, continuous parameter --- the JPA center frequency --- and both effects improve as the JPA bandwidth exceeds the bandwidth of the EMC and the EMC bandwidth exceeds the MO's center frequency.  

There are many ways to view this network.  For instance, this network may be interpreted as analogous to a traditional EMC, except with a user-controlled knob that modifies the coupling between the transmission line and resonator (the resonator center frequency may also be affected) \cite{Yin12}.  The specific network in Fig.~\ref{fig1}b can be viewed a form of microwave stub tuning \cite{Reed10,Yin12}, although the JPA bandwidth generalizes the usual model of a ``stub.''  This network may also be interpreted as a feedback control system: rather than make a room-temperature measurement of the JPA output, this signal is fed back directly to the EMC without leaving the cryostat.  In this work, the first interpretation aids physical intuition, but is not predictive.  The second has a physical intuition and is quantitatively predictive, but is limited to the specific wiring configuration used here.  The third, however, is also physically-intuitive and predictive, but may be applied quite generally.  

Although it is not our focus here, it is worth clarifying that coherent feedback approaches are distinct from measurement-based feedback approaches.  For example, in a measurement-based version of our system, coherent signals produced by the EMC and JPA would be measured by an incoherent controller (e.g. a computer), which would then synthesize new signals that act back on the MO \cite{Klec06,Gava12,LQG,Hame12}.  While a measurement-based approach can alter the response function of a MO, the system's effective EM linewidth is unchanged \cite{Klec06,Gava12}.  Here, the JPA's transformation of the microwave signal is ideally unitary (i.e. adds no noise).  This network is thus fully coherent and thus fundamentally different from any measurement-based control system \cite{G&J,Hinf,LQG,Goug10,Hame12,Wise94,Lloy00}.  For example, it has been noted that coherent feedback in the quantum regime can outperform even ideal measurement-based feedback when one considers the cost of the controller's action --- a measurement-based controller must act to cancel both environmental disturbances and quantum projection noise \cite{LQG,Hame12}.  At a more practical level, coherent control systems tend to be physically compact \cite{Kerc10,Hame12,Kerc12,Bott12,Wise94,Reed12,Legh12}.  We note that techniques borrowed from network systems theory are vital for tractable modeling of many coherent networks \cite{G&J,Goug10}.

Many of the demonstrations in the next section are analogous to the widely used electromechanical technique of sideband cooling \cite{Brag70,Teuf08,Teuf11,Mass12,Kipp08,Giga06,Arci06,Brow07,Chan11}.  The canonical sideband cooling system consists of an EMC, or some other electromechanical device, driven at a frequency below the resonator's center frequency.  Mechanical cooling via this method is usually explained in analogy to early ion trapping experiments \cite{Wils07,Marq07,Kipp08,Died89}.  A small portion of the low-entropy EM drive is inelastically scattered by a MO inside the resonator, with each scattered photon having gained or lost a phonon's worth of energy.  If the center frequency of the resonator is higher than the drive frequency, then more frequency up-converted photons are scattered than down-converted photons.  Consequently, the MO gives up energy on average to the EM field per scattering event, reducing its motion.  This phenomenon is usually visible through ``blue'' (up-converted) and ``red'' (down-converted) sidebands to the ``coupling tone'' (drive carrier) emitted by the resonator.  A cooling MO is indicated by more blue than red sideband power, less total sideband power, and sideband power spectral densities broader than the intrinsic MO linewidth.  To connect our coherent feedback network to sideband cooling, we point out that there are equivalent, coherent feedback interpretations of these phenomena \cite{Hame12,Bott12,Tsan10}.  These interpretations involve mechanically-modulated EM signals interfering with the coupling tone, say, at the input coupler.  Dependent on this interference, EM power is continuously added to or removed from the resonator in such a way to counteract mechanical motion.

\section{\label{sec:Results} Results}

\begin{figure*}[t]
\includegraphics[width=.75\textwidth]{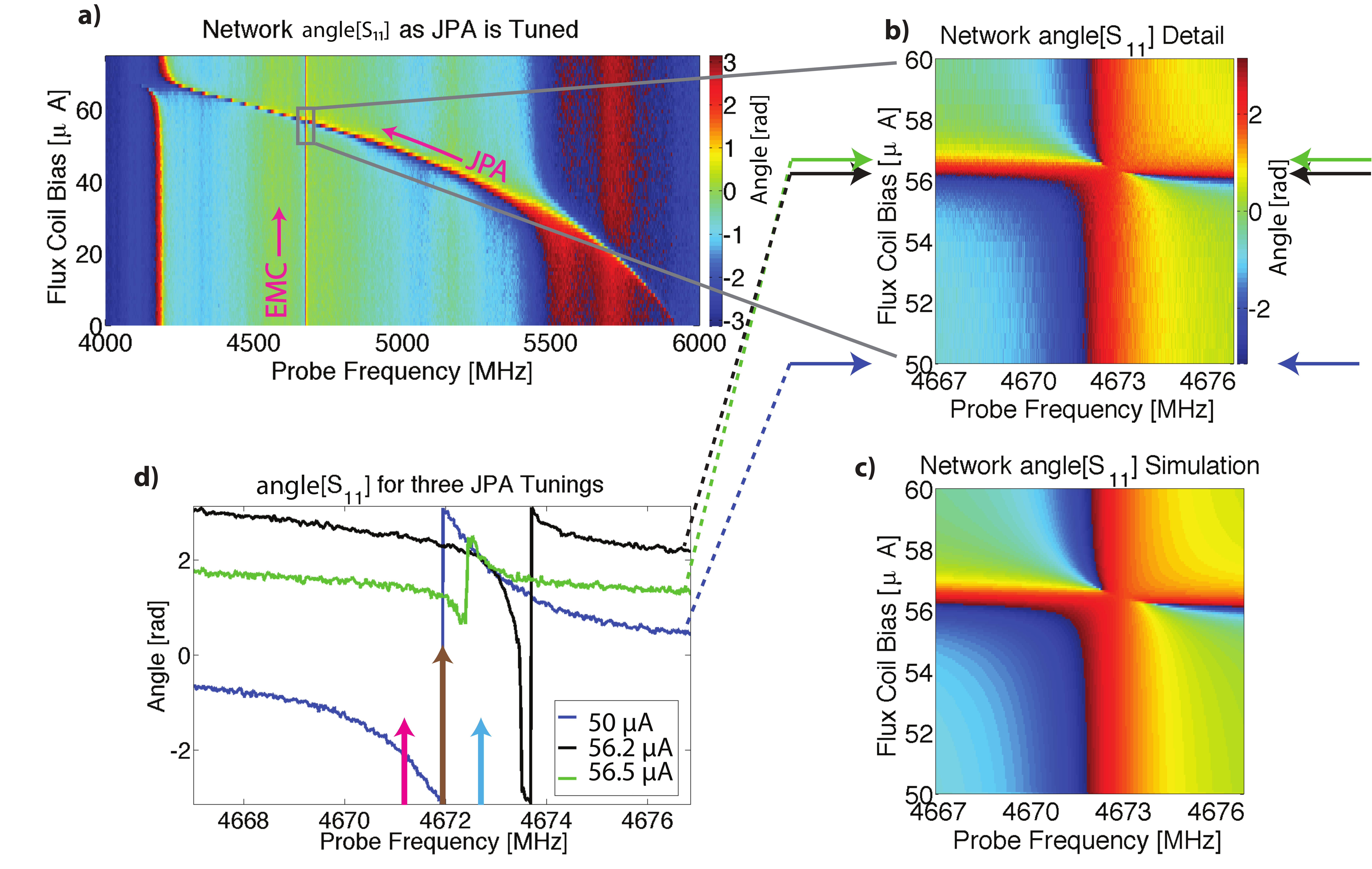}
\caption{\label{fig2} The network's microwave response. a) The network's 4-6 GHz $S_{11}$ phase response to microwave probes as the flux bias is tuned.  The JPA resonance shifts with flux and couples to EMC (4.7 GHz) and standing wave (4.2 \& 5.5 GHz) resonances as it passes through them.  b) Detail of the response around the EMC-JPA intersection.   c) Simulation of response detail using a linear control systems model.  d) Phase response line cuts at fixed, 50, 56.2 \& 56.5 $\mu$A flux coil biases, as marked by blue, black, and green arrows in b.  The JPA is far-detuned from the EMC at 50 $\mu$A coil bias, but near-detuned for the others.  The brown arrow represents an electromechanical coupling tone on-resonant with the 50 $\mu$A-biased network.  The blue and red arrows represent $\pm714$ kHz electromechanical sidebands (see Fig.~\ref{fig3}a). }
\vspace{-0.1in}
\end{figure*}


The network's response to microwave probe tones is depicted in Fig.~\ref{fig2}.  By monitoring the network's $S_{11}$ phase response (i.e. the phase with which microwave tones are reflected by the network) we see that the microwave response may be finely tuned using a static, magnetic flux bias applied to the JPA.  Viewing the response over a 2 GHz frequency range in Fig.~\ref{fig2}a, the JPA center frequency strongly varies with applied flux.  The other visible resonances, the narrow-band EMC and very broad-band resonances arising from standing waves resonances between the subcomponents, are only affected as the JPA tunes through them.  Focusing on the 10 MHz band around the intersection of the JPA and EMC resonances ($\approx56$ $\mu$A flux coil bias, Fig.~\ref{fig2}b), we see that as the flux increases and the JPA-like resonance approaches, the EMC-like resonance first broadens slightly, then abruptly narrows and shifts higher, then becomes under-coupled, and finally re-broadens and returns to its original center.  These effects indicate that the two device modes are coupling coherently and hybridizing.  

Microwave-related  parameters may be estimated given these 2 GHz and 10MHz probe ranges.  Using a linear control systems model built from interconnected device models (employing Matlab's Control Systems toolbox, see section~\ref{sec:Model}), we infer an EMC coupling linewidth of $\kappa_c=2\pi\times2.7$ MHz and internal loss rate $\kappa_l=2\pi\times0.1$ MHz; a JPA coupling linewidth of $\gamma=2\pi\times50$ MHz and internal loss rate of $\gamma_l=2\pi\times5$ MHz; and tee-to-EMC round trip phase shift of $\psi=-1.9$ rad and tee-to-JPA round trip phase phase shift of $\theta=-0.11$ rad.  These parameters are consistent with parameter estimates from the devices in isolation.  Although difficult to distinguish losses internal to these over-coupled devices from interconnection losses, a network model dominated by losses internal to the EMC and JPA is more consistent with observations than a network model dominated by interconnection losses.  Similarly, the tee is assumed to be ideal and properly terminated, reflecting 1/9$^{th}$ of the incident power with a $180^\circ$ phase shift and splitting the remaining power between the other two ports.  With these parameters, the linear control systems model accurately reproduces the network's microwave response as the flux varies, Fig.~\ref{fig2}c.

Only the EMC-like resonance (i.e. the resonance in the vicinity of 4.672 GHz) has significant coupling to the MO.  How this resonance responds to applied flux is more easily seen by taking a few, constant-flux line cuts through Fig.~\ref{fig2}b.  Fig.~\ref{fig2}d depicts three such line cuts.  With the JPA far-detuned (50 $\mu$A bias current), the microwave phase response indicates an overcoupled, $\kappa_n/2\pi=3.0$ MHz linewidth network resonance at 4.672 GHz.  Note that even with the JPA far detuned, the network causes the effective linewidth of the EMC-like resonance to differ from $\kappa_c+\kappa_l$, the linewidth of the bare EMC.  At different flux biasings, this resonance frequency and linewidth vary together, each by values of order the EMC's coupling linewidth.

\begin{figure}[b!]
\includegraphics[width=.5\textwidth]{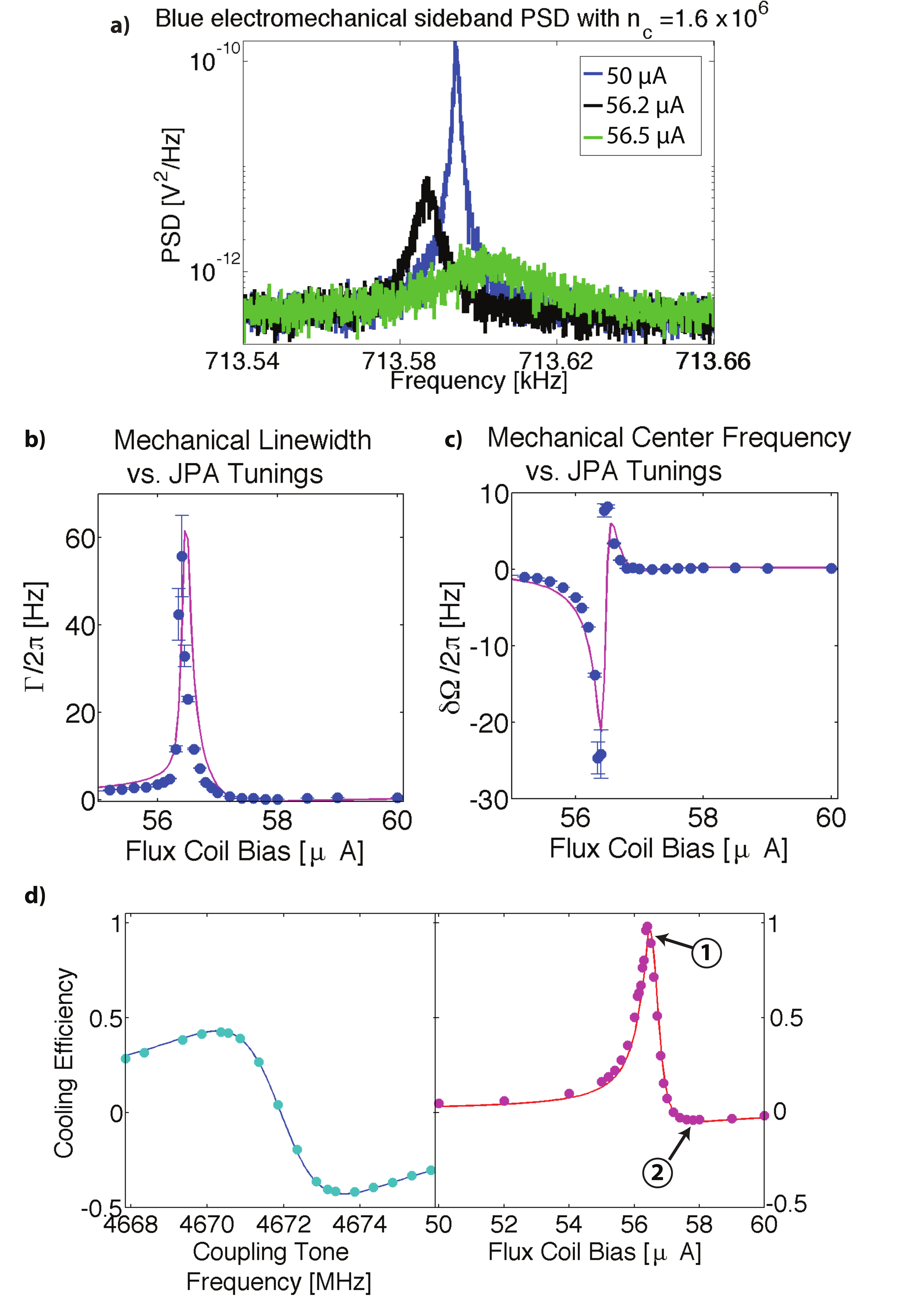}
\caption{\label{fig3} The network's mechanical response.  a)  Employing $n_c=1.6\times10^6$, the power spectral density of the blue, thermally-excited mechanical sideband is shown for three flux settings (same flux settings as in Fig.~\ref{fig2}d), keeping the coupling tone fixed at 4.672 GHz.  b) Points represent the observed linewidth of blue thermal sidebands as the JPA is tuned, employing a coupling tone at a fixed frequency and power such that $n_c=1.6\times10^6$.  The line is a theoretical prediction with no free parameters.  c) From the same experiment, the observed changes in the MO center frequency and theoretical prediction are shown.  d) Cooling efficiency (CE), as inferred from ring-down measurements and measured red and blue sideband powers.  Left, CE for different coupling tone frequencies when JPA is far-detuned.  Right, CE for a fixed coupling tone and different JPA tunings.  Lines are theoretical predictions.  Numbered call outs in right plot correspond to JPA tunings used in Fig.~\ref{fig4}a.}
\vspace{-0.1in}
\end{figure}

In Fig.~\ref{fig3}, we see that the dynamics of the MO also varies with applied flux.  Far-detuning the JPA once more (back to 50 $\mu$A flux coil bias) and driving the network with a strong coupling tone at the 4.672 GHz network resonance (and protecting the TKC with a cancellation tone through the directional coupler, Fig.~\ref{fig1}a), the thermal motion of the MO is visible via red and blue inelastically-scattered network output signals, i.e. sidebands.  The power spectral densities of these sidebands (Fig.~\ref{fig3}a) indicate a mechanical center frequency of $\Omega=2\pi\times713.6$ kHz and intrinsic damping rate of $\Gamma_0=2\pi\times0.81$ Hz.  Next, keeping the the coupling tone fixed at the same power and frequency, but varying the flux bias, the mechanical motion inferred from the sidebands varies.  In particular, as the network's resonance moves higher and its linewidth narrows in Fig.~\ref{fig2}d, in Fig.~\ref{fig3}a the apparent mechanical motion is damped more heavily (sideband linewidth becomes $\Gamma>\Gamma_0$), is cooled, and its center frequency moves ($\delta\Omega\neq0$).  These effects are analogous to sideband cooling in traditional electromechanical systems \cite{Kipp08}.  In our system, these effects may be interpreted as sideband cooling, or as the JPA acting as a coherent controller \cite{Hame12}, exerting different control laws for different amounts of applied flux.  

It is important to note that $n_c$, the number of photons in the EMC induced by the coupling tone, does not change as the TKC (the tunable component of the JPA) is tuned.  While the network's response to probes and mechanical sidebands varies with the TKC state, the coupling tone is canceled at the directional coupler before the TKC (Fig.~\ref{fig1}a) and is thus unaffected by the TKC's center frequency.  More precisely, the mean power in the coupling tone is dissipated in the terminated port in the directional coupler through interference with the cancellation tone, while information-carrying fluctuations about the mean power are not.  These fluctuations are fed back to the tee and EMC after reflecting off the TKC, while the mean coupling power experiences an entirely different network without TKC feedback.  Except for inducing $n_c$, this mean coupling power is unimportant and it carries no information.  The assumption of a constant $n_c$ is made throughout, and would have resulted in very inaccurate predictions in Figs.~\ref{fig3}b \& c (discussed next) if incorrect.   

While Fig.~\ref{fig3}a establishes the qualitative mechanical response, the rest of the figure considers it quantitatively.  Driving the network with a coupling tone fixed at a frequency just 13 kHz below the (4.672 GHz) network resonance with the JPA far-detuned and with a fixed power such that $n_c=1.6\times10^6$ is expected, Figs.~\ref{fig3}b \& c depict the linewidth and center frequency of the mechanical sidebands as a function of flux biasing.  Starting from a linewidth of $\Gamma\approx\Gamma_0$ with the JPA far-detuned, $\Gamma$ reaches a maximum of $2\pi\times56$ Hz, with a simultaneous frequency shift of $\delta\Omega=-2\pi\times24$ Hz, when the flux bias is 56.4 $\mu$A.  This flux bias corresponds to a JPA center frequency 3.4 MHz higher than the coupling tone, and an undercoupled network response with a $\approx300$ kHz internal loss rate.  If internal device losses were lower, this effective internal loss rate would be lower, and the maximum $\Gamma$ on Fig.~\ref{fig3}b would have been higher.  Because the near-resonant coupling tone has no significant effect on the mechanical state when the JPA is far-detuned, it is natural to interpret the JPA as ``controlling'' the MO as it is tuned through the EMC.  Using a control-systems model (here employing standard-issue Mathematica toolkits) and no free parameters, predictions for the expected mechanical linewidth and frequency shift as a function of flux bias are also included in Figs.~\ref{fig3} b \& c.

One metric for quantifying the efficiency with which the network controls the MO is the probability that a mechanically scattered photon dissipated by the network is a frequency up-converted photon minus the probability it is a down-converted photon.  This ``cooling efficiency'' (CE) is independent of $n_c$, but is coupling-frequency dependent and is positive when the network cools the MO (and is negative when it amplifies thermal motion).  The maximum CE obtainable over all coupling frequencies is $\Omega/\sqrt{\Omega^2+\kappa_n^2/4}$, where $\kappa_n$ is the total, EMC-like resonance linewidth.  In Fig.~\ref{fig3}d, left plot, we plot CE as a function of coupling tone frequency and with the JPA far-detuned.  These are measured by first inducing coherent oscillations in the MO well above their thermal occupation through amplitude modulation of the coupling tone at frequency $\Omega$, and then stopping the modulation and measuring the red and blue sidebands emitted by the network as the mechanical state re-equilibrates  \footnote{Besides representing a somewhat independent verification of the conclusions taken from the thermal data, these ``ring down'' measurements are more precise due to larger mechanical motion and motion that is phase-locked from trial to trial}.  Data is averaged over 25 trials.  The power emitted by both sidebands, $P_{red}$ and $P_{blue}$, decays exponentially (``rings down'') in time at rate $\Gamma$, while CE=$(P_{blue}-P_{red})/(P_{blue}+P_{red})$ is constant.  The extremum of CE are $\pm0.42$, implying a $\kappa_n/2\pi=3$ MHz-linewidth network coupled to the significantly slower $\Omega/2\pi=714$ kHz MO, an unresolved-sideband network.  Underlying this data is the expected CE versus coupling frequency, using independent network and MO calibrations.  Driving the network as in Figs.~\ref{fig3} b \& c and tuning the JPA, but now making ring down measurements, CE peaks at 0.98 at 56.4 $\mu$A.  A CE of 0.98 is only possible for networks with at most $\kappa_n/2\pi\approx300$ kHz, a resolved-sideband network and a 10-fold reduction in $\kappa_n$ consistent with the thermal data estimate above.  Such a reduction was apparent from microwave $S_{11}$ measurements in Fig.~\ref{fig2}, but CE represents an effective ``$S_{12}$'' measurement from the mechanical bath ``port''  to microwave output.  CE also reaches a minimum of -0.04 at 57.8 $\mu$A, indicating that the network resonance can also dip slightly below its far-detuned JPA resonance.  An independently calibrated control systems prediction using
\begin{equation}\label{eq:CE}
CE = \frac{\left|\Xi_{\{\mu_o\text{Th}_i\}}[j\Omega]\right|^2-\left|\Xi_{\{\mu_o^\ast\text{Th}_i\}}[j\Omega]\right|^2}{\left|\Xi_{\{\mu_o\text{Th}_i\}}[j\Omega]\right|^2+\left|\Xi_{\{\mu_o^\ast\text{Th}_i\}}[j\Omega]\right|^2}
\end{equation}
(see Eq.~\eqref{eq:TF}, section~\ref{sec:Model}) underlies the data.     

\begin{figure}[b!]
\includegraphics[width=.45\textwidth]{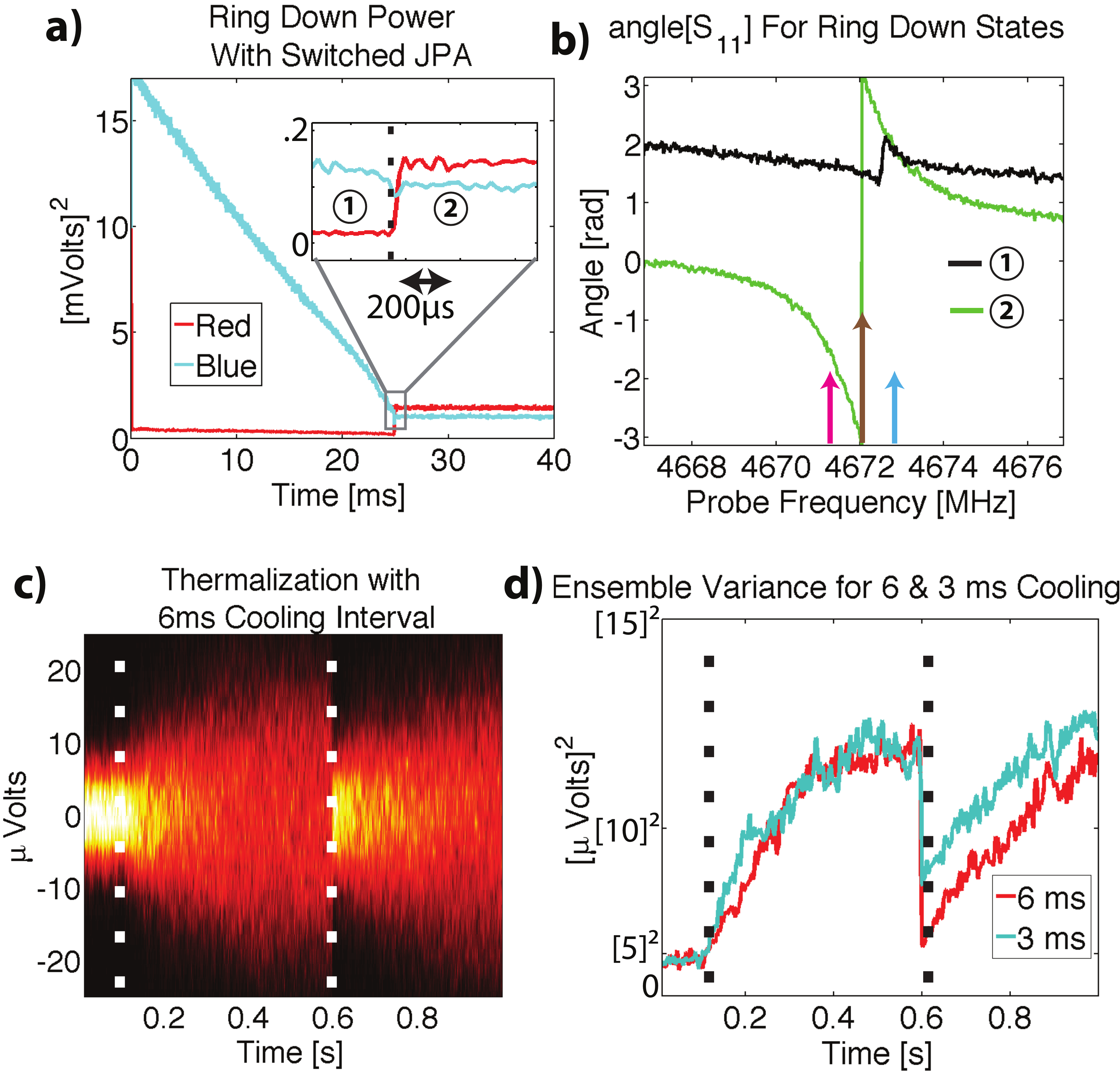}
\caption{\label{fig4} Dynamic JPA-control.  a) Red and blue sideband powers ringing down from a very high coherent amplitude.  From 0-25 ms (25-40 ms) the network is in state 1 (state 2), see Figs.~\ref{fig3}d \&~\ref{fig4}b.  Inset, detail of network transition, depicted with 25 kHz bandwidth.  b)  Microwave probe $S_{11}$ measurements for the network states employed in a, with arrows indicating coupling tone and sidebands.  c) Switched re-thermalization of a cooled MO.  At 100 ms, JPA is tuned from $\Gamma\gg\Gamma_{0}$ (cooling) state to $\Gamma=\Gamma_{0}$ (bath-thermalizing) state.  At 600 ms, JPA is switched to the cooling state for 6 ms only.  Plot represents the ensemble distribution of blue sideband amplitudes in time for 300 trials.  d) Sideband amplitude variance for 6 ms=$2/\Gamma$ and 3 ms=$1/\Gamma$ cooling-interval experiment ensembles.}
\vspace{-0.1in}
\end{figure}

Finally, we demonstrate that the state of the JPA, and thus the dynamics it effects, may be switched far faster than the MO can equilibrate.  While above we controlled the JPA state using static flux biasing, below we apply both static flux biasing and an additional microwave tone at a frequency 10MHz below the EMC-like resonance.  The amplitude of this new, ``control'' tone is switched dynamically and is not cancelled at the directional coupler.  At its strongest, the control tone is too weak and off-resonant with the EMC to affect the MO, but is sufficiently strong and resonant with the wider-band JPA to strongly shift its center frequency.  The rate with which the JPA may be switched in this manner should lie between the JPA linewidth (50 MHz) and the network's EM linewidth as a whole ($\geq$300 kHz ), orders of magnitude faster than the mechanical response rate (0.81 Hz intrinsic linewidth).  By performing similar ring down measurements starting from an even larger mechanical coherent amplitude (inducing $\sim$100 kHz pk-pk resonance shift in the EMC), this switched control may be observed with high visibility and bandwidth.  In Fig.~\ref{fig4}a, the power measured in the blue and red mechanical sidebands is depicted in time where, from 0 to 25 ms, the network is in a resolved-sideband and blue detuned state (state 1 in Fig.~\ref{fig4}b, with CE called out in Fig.~\ref{fig3}d), while from 25 to 40 ms, the network is in an unresolved-sideband and slightly red detuned state (state 2).  In the first segment, almost all the mechanically-scattered power is in the blue sideband, and the total power emitted decays in time, indicating MO cooling (the large amplitude motion employed here induces a variation in the EMC center frequency of order the state 1 linewidth.  As a consequence, the ring down power is more linear than exponential in time).  In the second segment, the two powers are roughly equal, with the red sidband slightly greater, and the total power hardly decays.  The inset depicts the transition between states 1 \& 2, through which the blue sideband  stays constant, but the red sideband power jumps by an order of magnitude.  This is only possible because state 2 has a much larger linewidth than 1 (Fig.~\ref{fig4}b).  The transition occurs at least as fast as the 25 kHz detection bandwidth, much faster than the mechanical decay rate in state 1 ($\Gamma_0\ll\Gamma\sim2\pi\times15$ Hz) or 2 ($\Gamma_0\gg\Gamma=2\pi\times0.1$ Hz).

Switched control over incoherent thermal motion may also be demonstrated, at the expense of reducing the detection bandwidth to 100 Hz (so that 10 min averaging times yield reasonable signal to noise ratios).  In Fig.~\ref{fig4}c \& d, the coupling tone and JPA are tuned such that the mechanics are cooled for several seconds with $\Gamma=2\pi\times56$ Hz (the most aggressive, resolved-sideband cooling configuration in Fig.~\ref{fig3}b).  At time 100 ms, the JPA is rapidly far-detuned from the EMC, so that the network is  on-resonance with the coupling tone and is in an unresolved sideband limit.  For 500 ms, the mechanical state then thermalizes with its bath at rate $\Gamma_0$.  After 500 ms, the JPA is briefly tuned to the cooling state for 6 ms, and then back to the thermalization state for a final 400 ms.  Fig.~\ref{fig4}c depicts the 300-trial ensemble distribution of continuously monitored blue sideband amplitudes.  Before 100 ms, the distribution is dominated by amplifier noise, but broadens by a factor of 3 during the thermalization periods.  The sideband amplitude is proportional to mechanical displacement (by a factor that differs for the cooling and thermalization states), and the broadening of the distribution is indicative of a warming MO (to $\sim750$ phonon occupation, or 26 mK, or (63 Hz)$^2$ variance in the EMC resonance, although precise knowledge of the thermal occupation is incidental to our purposes).  The position variance of the hot $\Gamma=2\pi\times56$ Hz MO is expected to drop by $e^{-2}$ in 6 ms, by $e^{-1}$ in 3 ms, and the variance of the cold, $\Gamma_0$-linewidth MO should rise to $1-e^{-1}$ of the bath equilibrium value in 200 ms.  Fig.~\ref{fig4}d depicts the ensemble variance of 6 \& 3 ms cooling-interval experiments and is consistent with these expectations.

\section{\label{sec:Model} Model}
Each subcomponent (the EMC, JPA and tee) is separably described by an I/O dynamical model, a familiar concept in quantum optics \cite{H&P,C&G,QN,G&J} and superconducting microwave systems \cite{Cler10,Kerc12}.  Moreover, despite fundamental nonlinearities, the device dynamics essential to the network's operation are accurately described by common linear approximations \cite{Wils07,Marq07}, as discussed more below.  It is well known that a network of linear, coherent I/O devices coupled to itinerant Gaussian fields may be modeled as a classical control system of interconnected, linear state space models \cite{Hinf,LQG,Goug10,Hame12,Iida11,Mabu08}.  The residual ``quantumness'' of these systems is captured by nonclassical ``noise'' driving the network inputs, but such quantum-level accuracy is not required in this work as the dynamics considered are classical.  Moreover, when vacuum fluctuations in the microwave fields are dynamically significant, each instance of microwave dissipation inside each device or interconnection between devices must be modeled using an additional I/O port pair \cite{G&J} (and a beamsplitter component interrupting lossy interconnections), but such accuracy is not required here.  In sum, the ``physics" of our model is captured by well-known I/O device models and cascaded interconnections \cite{QN,Wise94}, but our manipulation of these device models adopts techniques from network theory \cite{G&J,Goug10}.  To many readers, the least familiar aspect of the model is likely to be these manipulations, which are expected \cite{Hinf,Goug10,LQG,Hame12,Mabu08} (and partially tested \cite{Iida11}) to be the same in classical and quantum linear systems.

More precisely, the tee is modeled as a 6-input and 6-output (6-I/O) itinerant field scattering device (three physical ports, each of which is both a field input and output, each field described by two quadrature degrees of freedom) with no internal degrees of freedom.  The JPA is modeled as a 2-I/O, over-coupled linear resonator device (whose center frequency is a tunable parameter) with two internal degrees of freedom, representing the two quadratures of the mode inside the TKC \cite{QN,Cler10}.  And the EMC is a 4-I/O device, with two I/O ports connected to a microwave field (effectively at zero temperature) and two I/O ports connected to a thermal mechanical bath.  The EMC also contains four internal degrees of freedom, representing the quadratures of coupled microwave and mechanical modes \cite{Hame12,Bott12,Tsan10}.  Consequently, a linear network model representing the schematic depicted in Fig.~\ref{fig1}b consists of ten coupled linear equations of motion, with six internal degrees of freedom, and two microwave and two mechanical bath I/O port pairs.  Despite this complexity, standard control systems software toolkits make the construction of network dynamical models intuitive and efficient linear system theories may be applied to their analysis.  We now describe the model construction in more detail. 

While the electromechanical interaction is fundamentally nonlinear, the dynamics we consider may be modeled by linearizing the coupling of the microwave and mechanical modes about the large microwave coherent state induced in the resonator by the coupling tone \cite{Wils07,Marq07}, an approximation that is nearly ubiquitous in electromechanics.  In a frame in which the EM degrees of freedom are rotating with the coupling tone's carrier frequency, the Hamiltonian that describes the internal dynamics of the EMC in the linearized approximation is ($\hbar=1$)
\begin{equation}
H_{M}=\Omega a_1^\dag a_1+\Delta a_2^\dag a_2 + g(a_1+a_1^\dag)(a_2+a_2^\dag)
\end{equation}
where $a_1$ is the annihilation operator for the mechanical mode and $a_2$ the annihilation operator for fluctuations in the microwave mode about the large coherent state ($a_{1,2}^\dag$ the creation operators).  Additionally, $\Omega$ is the center frequency of the MO, $\Delta=\omega_r-\omega_c$ is the frequency detuning between the resonator center frequency ($\omega_r$) and the carrier of the coupling tone ($\omega_c$), and $g=g_0\sqrt{n_c}$, where $n_c$ is the number of intra-resonator photons induced by the coupling tone and $g_0$ is the fundamental coupling rate between photons and phonons.  

In the usual I/O formulation for an EMC, the Heisenberg equations of motion may be written as \cite{Hame12,Bott12}
\begin{widetext}\begin{eqnarray}\label{eq:EMC}
\frac{d}{dt} \left[\begin{array}{c}a_1\\ a_2\\ a_1^\dag\\ a_2^\dag\end{array}\right] &=&\left[\begin{array}{cccc} 
j\Omega-\frac{\Gamma_0}{2} & jg & 0 & jg\\
jg & j\Delta - \frac{\kappa_l+\kappa_c}{2} & jg & 0\\
0 & -jg & -j\Omega-\frac{\Gamma_0}{2} & -jg\\
-jg & 0 & -jg & -j\Delta - \frac{\kappa_l+\kappa_c}{2}\end{array}\right]\left[\begin{array}{c}a_1\\ a_2\\ a_1^\dag\\ a_2^\dag\end{array}\right]+\left[\begin{array}{cccc}-j\sqrt{\Gamma_0}&0&0&0\\ 0&-j\sqrt{\kappa_c}&0&0\\0&0&j\sqrt{\Gamma_0}&0\\0&0&0&j\sqrt{\kappa_c}\end{array}\right]\left[\begin{array}{c}b_{Th,in}\\ b_{E,in}\\ b_{Th,in}^\dag\\ b_{E,in}^\dag\end{array}\right]\nonumber\\
\left[\begin{array}{c}b_{Th,out}\\ b_{E,out}\\ b_{Th,out}^\dag\\ b_{E,out}^\dag\end{array}\right]&=&\left[\begin{array}{cccc}j\sqrt{\Gamma_0}&0&0&0\\ 0&j\sqrt{\kappa_c}&0&0\\0&0&-j\sqrt{\Gamma_0}&0\\0&0&0&-j\sqrt{\kappa_c}\end{array}\right] \left[\begin{array}{c}a_1\\ a_2\\ a_1^\dag\\ a_2^\dag\end{array}\right] +\left[\begin{array}{cccc}-1&0&0&0\\ 0&-1&0&0\\0&0&-1&0\\0&0&0&-1\end{array}\right]\left[\begin{array}{c}b_{Th,in}\\ b_{E,in}\\ b_{Th,in}^\dag\\ b_{E,in}^\dag\end{array}\right]
\end{eqnarray}\end{widetext}
where, $b_{Th,in}$ and $b_{E,in}$ ($b_{Th,out}$ and $b_{E,out}$) are standard mathematical objects in I/O theory that are functions of time $t$ and may be roughly considered annihilation operators on the infinitesimal segment of free field incident on (leaving from) the device at time $t$ \cite{Goug10,C&G,H&P,QN}.  Specifically, $b_{Th,in/out}$ and $b_{E,in/out}$ are associated with the fields that drive the system through the thermal mechanical port and the EMC's microwave port, respectively.  As with $a_2$, $b_{E,in/out}$ is associated with the microwave field fluctuations about the coupling tone amplitude.  Moreover, $\Gamma_0$ is the intrinsic mechanical decay rate, $\kappa_c$ ($\kappa_l$) is the EMC coupling (internal) decay rate, and $j$ is the imaginary number with the electrical engineering sign convention \footnote{While physicists tend to define a Fourier transform from the time to frequency domain as $f[w]=\int_{-\infty}^{\infty}f(t)e^{i\omega t}dt$ where $i$ is the imaginary number, in electrical engineering contexts it is more common to find the same transform defined as $f[w]=\int_{-\infty}^{\infty}f(t)e^{-j\omega t}dt$ where $j$ is the imaginary number.}.

Eqs.~\eqref{eq:EMC} are unwieldy, but they also contain relatively few free parameters.  Linear, coherent I/O systems feature such restrictions in general, which may be traced back to fundamentally unitary dynamics (e.g. I/O theory in the quantum stochastic differential equation formulation \cite{C&G,H&P})) and are the primary distinction between coherent and more general linear I/O systems \cite{Goug10,Hame12}.  Rather than specify the fully reduced form of Eqs.~\eqref{eq:EMC} \cite{Goug10, Hame12}, we simply note that they may be written as
\begin{eqnarray}\label{eq:EMC2}
\frac{d}{dt}{\breve{a}}_{} &=& A_{M}\,\breve{a}_{}+B_{M}\,\breve{b}_{in}\nonumber\\
\breve{b}_{out} &=& C_{M}\,\breve{a}+D_{M}\,\breve{b}_{in}
\end{eqnarray}
where $\breve{a}_{} = \left[a_1,a_2,a_1^\dag,a_2^\dag\right]^T$ ($T$ indicating transpose) is a vector of operators, $\breve{b}_{in}=\left[b_{Th,in},b_{E,in},b_{Th,in}^\dag,b_{E,in}^\dag\right]^T$, and similarly for $\breve{b}_{out}$.  The matrices $A_{M},\,B_{M},\,C_{M},\,D_{M}$ define the device dynamics.

Eqs.~\ref{eq:EMC2} are directly relatable to a common representation of a classical linear system known as a state space model.  Emulating this representation even more fully, we can represent Eqs.~\ref{eq:EMC2} using the mathematical object $\mathbb{M}$, which is defined by $A_M$, $B_M$, $C_M$ and $D_M$ and notated \cite{AM,Goug10}
\begin{equation}\label{eq:EMCss}
\mathbb{M}=\left[\begin{array}{c|c}A_M&B_M\\\hline
C_M&D_M\end{array}\right]
\end{equation}  
(each matrix and its location in the above array implies Eqs.~\ref{eq:EMC2}).  We can similarly generate state space model representations for the JPA and tee devices \cite{Goug10}; call them $\mathbb{J}$ and $\mathbb{T}$, respectively.

\begin{figure}[b!]
\includegraphics[width=.5\textwidth]{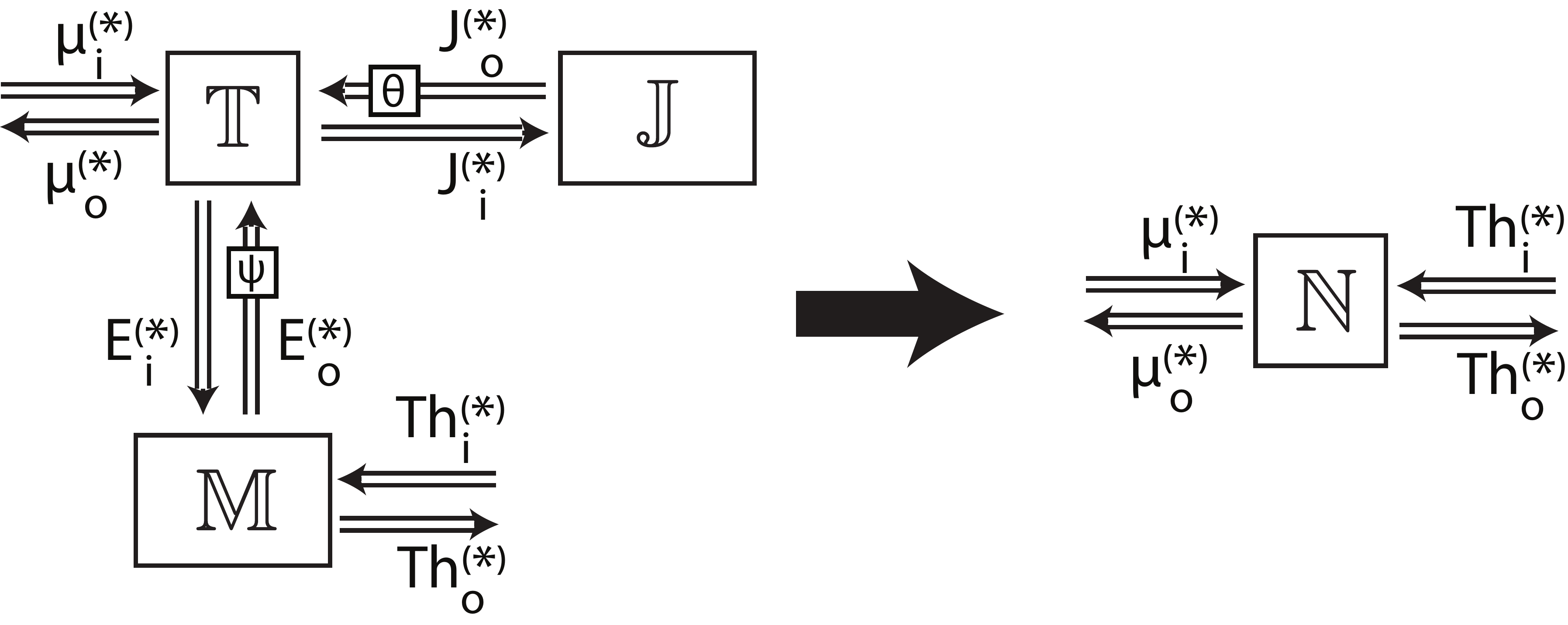}
\caption{\label{figFB} Control system schematic of the network depicted in Fig.~\ref{fig1}b.  The tee, EMC and JPA components are represented as multi-input multi-output (MIMO) linear state space models that exchange signals continuously.  The interconnections are labeled here such that, e.g. $E_o^{(\ast)}$ stands for the two free field channels labeled $E_o$ and $E_o^\ast$, which are explained in the text.  Transmission line delays may be modeled here as static phase shifts $\psi$ and $\theta$.  Standard software toolkits are available that reduce the network on the left to a single MIMO state space model $\mathbb{N}$.  This approach is compatible with both quantum and classical investigations.}
\vspace{-0.1in}
\end{figure}

We now come to the problem of constructing the network model.  As discussed in Refs.~\cite{Goug10,G&J,Hame12} and elsewhere, the formal relation between linear quantum I/O device models and classical state space models means that procedures for constructing dynamical models of linear quantum networks are analogous to those for classical linear networks.  Through these methods, we may derive a new state space model $\mathbb{N}$ that describes the network as a whole.  In the remainder of this section, it will be ambiguous whether we are constructing a quantum or coherent classical model.  It is only when we neglect the effects of quantum fluctuation in the free fields when we compare our model to experiment that the approach becomes a classical approximation \cite{Mabu08,Kerc12}.

The network model we need is depicted schematically in Fig.~\ref{figFB}, as inspired by Fig.~\ref{fig1}b, where the EMC and JPA each exchange input and output signals with two ports of the tee.  The ``$\psi$'' and ``$\theta$'' blocks represent the phase shifts accumulated just from transmission line delays around the tee-and-EMC and tee-and-JPA network loops, respectively \footnote{A more physical model would not lump these phase delays only on the input side of the tee, but half on each input and output.  Up to unimportant phase shifts to the network internal variables, the two modeling approaches yield identical predictions.  For the dynamics considered here, transmission line delays may be approximated by static, dispersionless phase shifts between the subcomponents.}.  We thus define our network by defining the state space model objects $\mathbb{T}$, $\mathbb{M}$, and $\mathbb{J}$, and labeling the channels associated with each component's input and output fields according to the schematic Fig.~\ref{figFB}.  For instance, the label ``$E_o$''  (the label ``$E_o^\ast$'') in Fig.~\ref{figFB} is applied to both the channel through which the $\mathbb{M}$ output field $b_{E,out}$ ($b_{E,out}^\dag$) ``flows'' and the appropriate channel through which one of the $\mathbb{T}$ inputs is driven.  The label ``$\mu_i$'' is applied to the channel through which the (heretofore unmentioned) $b_{\mu,in}$ field drives a $\mathbb{T}$ input associated with the network's microwave input, etc.  

While algorithmic, constructing $\mathbb{N}$ using a linear systems approach is extremely tedious.  However, software toolboxes are available in, for example, Matlab and Mathematica that completely automate the procedure.  This basic fact is under-recognized by the physics community and deserves explicit emphasis: after defining the three state space models and labeling the component input and output channels according to Fig.~\ref{figFB}, the entire state space model $\mathbb{N}$ may be obtained in Matlab using the single command {\tt N = connect(T,M,J)}.  (other software that can deal with nonlinear quantum I/O networks has also been developed \cite{Teza11,Kerc12})  Using such methods, these control systems toolkits determine an effective network state space model defined by matrices $A_N$ (dimensions $6\times6$), $B_N$ ($6\times4$), $C_N$ ($4\times6$), and $D_N$ ($4\times4$), 
\begin{equation}\label{eq:Nss}
\mathbb{N}=\left[\begin{array}{c|c}A_N&B_N\\\hline
C_N&D_N\end{array}\right],
\end{equation}  
which are too unwieldy to write out in general here.  

It is useful, though, to consider the network's equation of motion for just $a_2$, the annihilation operator for the microwave fluctuations in the EMC.  (The network modifies the dynamics of the MO only inasmuch as $a_2$ and $a_2^\dag$ are modified)  For the EMC in isolation, and taking $g=0$ for clarity, the equation of motion for this mode is (Eq.~\eqref{eq:EMC})
\begin{equation}\label{eq:a2}
\frac{d}{dt}a_2=\left(j\Delta-\frac{\kappa_t}{2}\right)a_2-j\sqrt{\kappa_c}b_{E,in}
\end{equation}
where $\kappa_t=\kappa_l+\kappa_c$.  In the network $\mathbb{N}$, taking $g=0$ and in the limit that the JPA has no internal loss and couples much more strongly to transmission lines than the EMC does, we find that
\begin{equation}\label{eq:a2eff}
\frac{d}{dt}a_2=\left(j\Delta^\prime_{\theta^\prime}-\frac{\kappa_n}{2}\right)a_2+e^{-j(\lambda+\frac12\psi)}\sqrt{\kappa_c^\prime}b_{\mu,in}
\end{equation}  
where $\kappa_n=\kappa_l+\kappa_c^\prime$ and
\begin{eqnarray}\label{eq:primes}
\kappa_c^\prime &=& \kappa_c\frac{\cos^2\left(\frac{\theta^\prime}{2}\right)}{\cos^2\left(\frac{\theta^\prime}{2}\right)\sin^2\left(\frac{\psi}{2}\right)+\cos^2\left(\frac{\psi+\theta^\prime}{2}\right)}\nonumber\\
\Delta^\prime &=& \Delta - \kappa_c\frac14\frac{\cos^2\left(\frac{\theta^\prime}{2}\right)\sin(\psi)-\sin(\theta^\prime+\psi)}{\cos^2\left(\frac{\theta^\prime}{2}\right)\sin^2\left(\frac{\psi}{2}\right)+\cos^2\left(\frac{\psi+\theta^\prime}{2}\right)}\nonumber\\
\lambda &=& \arctan\left(\cot\left(\frac{\psi}{2}\right)-\tan\left(\frac{\theta^\prime}{2}\right)\right)
\end{eqnarray}
where $\theta^\prime$ is the total phase shift acquired by a narrow-band signal that makes a round trip from the tee to JPA and back (i.e. is determined by the length of that interconnection and the center frequency of the JPA).  

From the correspondence between Eqs.~\eqref{eq:a2} \&~\eqref{eq:a2eff}, we see that the EMC resonator response still acts like a single-mode resonator in this limit, except one that is now driven by the inputs ``$\mu^{(\ast)}_{i}$'' and has a new detuning from the coupling tone ($\Delta^\prime$), a new decay rate ($\kappa_n$), and a new phase shift between inputs and outputs (controlled by $\lambda$) that are all controlled by $\theta^\prime$, which is in turn controlled by the JPA center frequency.  New effective system parameters, such as new effective decay rates and detunings, is a primary reason for constructing coherent feedback networks in general \cite{Kerc10,Goug10,Hame12,Kerc12}.  In this case, these modifications quantitatively express the physical mechanisms qualitatively described in section~\ref{sec:Network}.  For example, the effective coupling rate $\kappa_c^\prime$ comes from the interference between signals emitted by $\mathbb{M}$ that are passed to the network output directly and those that reflect off $\mathbb{J}$ before being passed to the same output port.  The effective detuning, $\Delta^\prime$, represents the interference between fields inside $\mathbb{M}$ and those that exit $\mathbb{M}$, but are fed back into $\mathbb{M}$  by the network.  In general, and especially when intra-JPA loss is not negligible, these same qualitative effects still hold, although they lose their quantitative accuracy, in which case the full network model $\mathbb{N}$ is needed.

 If losses were negligible and the JPA bandwidth effectively infinite, the effective microwave parameters given in Eq. (\ref{eq:primes}) could be substituted immediately into a traditional analysis of an electromechanical circuit \cite{Wils07,Marq07,Kipp08,Teuf11}, whose physical interpretations are discussed in section~\ref{sec:Network}.  Eq. (\ref{eq:a2eff}) could also be extended to incorporate non-idealities with yet more complicated expressions, but this is beside the point.  The modularity of the devices that make up the network permits us to use modular network techniques to construct a dynamical model and make predictions using general and efficient algorithms.  Analytic expressions of the full equations of motion are readily available, but they complicate matters unnecessarily here, and they are efficiently reproduced through the construction outlined above.  Furthermore, we argue that such an approach is often efficient, useful, and appropriate as electromechanics and quantum engineering in general begins to move beyond the physics of individual devices.
 
 In section~\ref{sec:Results}, we compare the network model $\mathbb{N}$'s steady state output predictions to data.  This is most naturally done through a Laplace transform of the field and internal variables such that, e.g. 
\begin{equation}
b_{\mu,in}[s] = \int_0^\infty e^{-st}b_{\mu,in}dt
\end{equation}
where $b_{\mu,in}$ (as above) is a function of time.  In the Laplace domain, it is appropriate to write the network equations of motion in terms of susceptibilities and transfer functions such that \cite{Goug10}
\begin{eqnarray}\label{eq:TF}
\breve{a}^\prime[s]&=&(sI-A_N)^{-1}B_N\breve{b}^\prime_{in}[s]\equiv\chi[s]\breve{b}^\prime_{in}[s]\nonumber\\
\breve{b}^\prime_{out}[s]&=&\left(C_N(sI-A_N)^{-1}B_N+D_N\right)\breve{b}^\prime_{in}[s]\equiv\Xi[s]\breve{b}^\prime_{in}[s]\nonumber\\
\end{eqnarray}
where $\breve{a}^\prime=[a_1,\,a_2,\,a_3,\,a_1^\dag,\,a_2^\dag,\,a_3^\dag]^T$ where $a_3$ is the annihilation operator on the internal JPA mode, $\breve{b}^\prime_{in/out}=[b_{Th,in/out},\,b_{\mu,in/out},\,b_{Th,in/out}^\dag,\,b_{\mu,in/out}^\dag]^T$, and $I$ is a $6\times6$ identity matrix.  While manually calculating the multi-input multi-output (MIMO) transfer function $\Xi[s]$ would be prohibitively tedious, again standard software toolboxes completely automate the procedure; in Matlab, given a network state space model {\tt N}, the MIMO transfer function representation is obtained using the single command {\tt Xi = tf(N)}.

The transfer function $\Xi[s]$ is applied several times in section~\ref{sec:Results} to make predictions about the signals emitted by the network.  For instance, the network's phase response to microwave probes of varying frequency is simulated in Fig.~\ref{fig2}c by calculating the phase angle of the $\{\mu_\text{o},\mu_\text{i}\}$ matrix element of $\Xi[s]$, $\text{angle}\left[\Xi_{\{\mu_o \mu_i\}}[s]\right]$, as $s$ runs along the imaginary axis.  Similarly, the absolute value squared of the $\{\mu_\text{o},\text{Th}_\text{i}\}$ matrix element of $\Xi[j\Omega]$, $\left|\Xi_{\{\mu_o\text{Th}_i\}}[j\Omega]\right|^2$, represents the gain with which $\Omega$-energy excitations in the thermal bath induce $\Omega$-energy excitations in the microwave network output (i.e. induce blue sideband power in the network output), through effective $\left(b_{\mu,out}^\dag b_{Th,in}+h.c.\right)$-type interactions (used in Eq.~\eqref{eq:CE}).  Conversely, $\left|\Xi_{\{\mu_o^\ast\text{Th}_i\}}[j\Omega]\right|^2$ represents the gain of thermal bath-induced $\Omega$-energy de-excitations in the microwave network output (i.e. induction of red sideband power in the network output), through effective $\left(b_{\mu,out} b_{Th,in}+h.c.\right)$-type interactions.  Similarly, 
\begin{equation}
\operatorname*{arg\,max}_{\delta\Omega}\left|\Xi_{\{\mu_o\text{Th}_i\}}[j(\Omega+\delta\Omega)]\right|^2
\end{equation}
may be used to predict microwave-induced shifts in the mechanical center frequency \cite{Brow07,Teuf08} (used in Fig.~\ref{fig3}c).  Finally, an expression related to $\chi[j\Omega]$, the $\{\mu_o,a_2\}$ matrix element of
\begin{equation}
C_N(j\Omega I-A_N)^{-1}jg
\end{equation}
 may be used to predict the rate with which MO excitations decay out the network's blue sideband, potentially enhancing the total MO decay rate, and similar expressions may be used to predict red sideband decay and losses dissipated in the network internally (used in Fig.~\ref{fig3}b).  Thus, the functions $\Xi[s]$ and $\chi[s]$ are extremely convenient for modeling steady state network dynamics.  And while the matrix of expressions represented by $\Xi[s]$ is difficult for a human to parse, for instance, the mathematical object $\Xi[s]$ is easily manipulated algebraically and computationally.

\section{\label{sec:Con} Conclusion}
We have demonstrated a small coherent feedback network of modular superconducting microwave devices that provides a type of dynamical flexibility previously unavailable to electromechanics.  The network has at least three natural interpretations: a dynamically-tunable input coupler, a tunable microwave stub \cite{Reed10,Yin12}, or a gain-1 amplifier that feeds back coherent signals \cite{Hame12}.  Despite the simplicity of the components and construction (coaxial cabling between pre-existing and familiar superconducting devices), the network is too complex to be modeled using traditional electromechanical techniques (e.g., such approaches assume a single resonator mode, while our network features coupled resonators) \cite{Wils07,Marq07,Kipp08}.  However, it is efficiently and intuitively modeled using coherent network techniques \cite{G&J,Goug10,Hinf,LQG,Hame12}, an unconventional approach that will find increasing utility as electromechanics (or quantum engineering in general) begins to move beyond the physics of individual devices.  Although we only demonstrate this network's operation in the classical regime, on the basis of the well-known connections between classical and quantum dynamics in linear coherent systems \cite{G&J,Goug10,Hinf,LQG,Hame12,Mabu08}, we expect that the essential mechanisms of the network should work analogously in the presence of unambiguously quantum fields and states.  We have not yet probed the quantum regime because of our use of a HEMT (non-quantum limited) amplifier for readout and the dynamical richness to be explored first in the classical regime.

There are several worthwhile directions for future investigations with this same network.  First, the TKC may be pumped by a third microwave tone to add a parametric gain-and-squeezing element \cite{Cast08} (or a more general Kerr nonlinearity \cite{Kerc12}) into the feedback and read-out dynamics.  Recent theoretical work suggests that quantum coherent feedback to an electromechanical system from a parametric gain controller can outperform any type of ideal measurement-based feedback or passive coherent controller \cite{LQG,Hame12}.  Second, the ability to dynamically and continuously modulate the network's coupling to its I/O port could be leveraged to shape the waveform of signals read into and out of the EMC.  This capability could facilitate high-fidelity coherent state transfer between the MO and arbitrary coherent devices either ``up-'' or ``down-stream'' from the network \cite{Cira97,Palo13,Yin12}.  And finally, the model could be considered from the perspective of well-developed theories of classical optimal and robust control \cite{AM,Hinf,LQG,Hame12}.  Such investigations are likely to yield recommendations for a more precisely-controlled network construction.  In particular, different applications are helped and hindered by different round trip phase shifts over the EMC and TKC network branches.  While these phases were uncontrolled by the use of bulky cable interconnections in this work, a more integrated network could be constructed with much better precision. 

We acknowledge partial support from the DARPA QuEST program, the DARPA ORCHID program, and from the NSF Physics Frontier Center at JILA.  JK acknowledges the NRC for financial support.


\begin{thebibliography}{99}

\bibitem{G&J} J. Gough and M. R. James, {\it The Series Product and its Application to Quantum Feedforward and Feedback Networks}, IEEE Trans. Auto. Control.~{\bf 54}, 2530 (2009); {\it Quantum Feedback Networks: Hamiltonian Formulation}, Commun. in Math. Phys.~{\bf 287}, 1109 (2009).

\bibitem{AM} K. J. \r{A}str\"{o}m and R. M. Murray, {\it Feedback Systems: An Introduction for Scientists and Engineers} (Princeton University Press, Princeton, New Jersey, 2008).
\bibitem{WM} H. M. Wiseman and G. J. Milburn, {\it Quantum Measurement and Control} (Cambridge University Press, Cambridge, England, 2009).

\bibitem{Kipp08} T.~J.~Kippenberg and K.~J.~Vahala, {\it Cavity Optomechanics: Back-Action and the Mesoscale}, Science {\bf 321}, 1172 (2008).
\bibitem{OCon10} A.~D.~O'Connell {\it et al.}, {\it Quantum ground state and single-phonon control of a mechanical resonator}, Nature (London) {\bf 464}, 697 (2010).
\bibitem{Teuf11} J.~D.~Teufel, T.~Donner, D.~Li, J.~W.~Harlow, M.~S.~Allman, K.~Cicak, A.~J.~Sirois, J.~D.~Whittaker, K.~W.~Lehnert, and R.~W.~Simmonds, {\it Sideband Cooling Micromechanical Motion to the Quantum Ground State}, Nature (London) {\bf 475} 359, (2011).
\bibitem{Chan11} J.~Chan, T.~P.~Mayer~Alegre, A.~H.~Safavi-Naeini, J.~T.~Hill, A.~Krause, S.~Gr\"{o}blacher, M.~Aspelmeyer, and O.~Painter, {\it Laser Cooling of a Nanomechanical Oscillator into its
Quantum Ground State}, Nature (London) {\bf 478}, 89 (2011).
\bibitem{Safa12} A.~H.~Safavi-Naeini, J.~Chan, J.~T.~Hill, T.~P.~Mayer~Alegre, A.~Krause, and O.~Painter, {\it Observation of Quantum Motion of a Nanomechanical Resonator}, Phys. Rev. Lett. {\bf 108}, 033602 (2012).
\bibitem{Purd13} T.~P.~Purdy, R.~W.~Peterson and C.~A.~Regal, {\it Observation of Radiation Pressure Shot Noise on a Macroscopic Object}, Science {\bf 339}, 801 (2013).
\bibitem{Palo13} T.~A.~Palomaki, J.~W.~Harlow, J.~D.~Teufel, R.~W.~Simmonds, and K.~W.~Lehnert, {\it Coherent State Transfer Between Itinerant Microwave Fields and a Mechanical Oscillator}, Nature (London) {\bf 495}, 210 (2013).

\bibitem{Teuf08} J.~D.~Teufel, J.~W.~Harlow, C.~A.~Regal and K.~W.~Lehnert, {\it Dynamical Backaction of Microwave Fields on a Nanomechanical Oscillator}, Phys. Rev. Lett. {\bf 101}, 197203 (2008).
\bibitem{LaHa09} M.~D.~LaHaye, J.~Suh, P.~M.~Echternach, K.~C.~Schwab, and M.~L.~Roukes, {\it Nanomechanical Measurements of a Superconducting Qubit}, Nature (London) {\bf 459}, 960 (2009).
\bibitem{Teuf09} J.~D.~Teufel, T.~Donner, M.~A.~Castellanos-Beltran, J.~W.~Harlow, and K.~W.~Lehnert, {\it Nanomechanical Motion Measured with Imprecision Below that at the Standard Quantum Limit}, Nature Nanotechnol. {\bf 4}, 820 (2009).
\bibitem{Hertz10} J.~B.~Hertzberg, T.~Rocheleau, T.~Ndukum, M.~Savva, A.~A.~Clerk, and K.~C.~Schwab, {\it Back-Action-Evading Measurements of Nanomechanical Motion}, Nature Phys. {\bf 6}, 213 (2010).
\bibitem{Mass12} F.~Massel, S.~U.~Cho, J.-M.~Pirkkalainen, P.~J.~Hakonen, T.~T.~Heikkil\"{a}, and M.~A.~Sillanp\"{a}\"{a}, {\it Multimode Circuit Optomechanics Near the Quantum Limit}, Nature Comm. {\bf 3}, 987 (2012).
\bibitem{Rega11} C.~A.~Regal and K.~W.~Lehnert, {\it From Cavity Electromechanics to Cavity Optomechanics}, Journal of Physics: Conference Series {\bf 264}, 012025 (2011).
\bibitem{Andr13} R.~W.~Andrews {\it et al.}, {\it in preparation.}

\bibitem{Cast08} M.~A.~Castellanos-Beltran, K.~D.~Irwin, G.~C.~Hilton, L.~R.~Vale, and K.~W.~Lehnert, {\it Amplification and Squeezing of Quantum Noise with a Tunable Josephson Metamaterial}, Nature Phys. {\bf 4}, 929 (2008).
\bibitem{Kerc12} J. Kerckhoff and K. W. Lehnert, {\it Superconducting Microwave Multivibrator Produced by Coherent Feedback}, Phys. Rev. Lett. {\bf 109}, 153602 (2012). 

\bibitem{Wils07} I.~Wilson-Rae, N.~Nooshi, W.~Zwerger, and T.~J.~Kippenberg, {\it Theory of Ground State Cooling of a Mechanical Oscillator Using Dynamical Backaction}, Phys. Rev. Lett. {\bf 99}, 093901 (2007).
\bibitem{Marq07} F.~Marquardt, J.~P.~Chen, A.~A.~Clerk, and S.~M.~Girvin, {\it Quantum Theory of Cavity-Assisted Sideband Cooling of Mechanical Motion}, Phys. Rev. Lett. {\bf 99}, 093902 (2007).

\bibitem{Goug10} J.~E.~Gough, M.~R.~James, and H.~I.~Nurdin, {\it Squeezing Components in Linear Quantum Feedback Networks}, Phys. Rev. A~{\bf 81}, 023804 (2010). 
\bibitem{Hinf} M.~R.~James, H.~I.~Nurdin, and I.~R.~Petersen, {\it H$^{\infty}$ Control of Linear Quantum Stochastic Systems}, IEEE Trans. Auto. Control. {53}, 1787 (2008).
\bibitem{LQG} H.~I.~Nurdin, M.~R.~James, and I.~R.~Petersen, {\it Coherent quantum LQG control}, Automatica ~{\bf 45}, 1837 (2009).
\bibitem{Hame12} R.~Hamerly and H.~Mabuchi, {\it Advantages of Coherent Feedback for Cooling Quantum Oscillators}, Phys. Rev. Lett. {\bf 109}, 173602 (2012).
\bibitem{Mabu08} H.~Mabuchi, {\it Coherent-feedback Quantum Control with a Dynamic Compensator}, Phys. Rev. A~{\bf 78}, 032323 (2008).
\bibitem{Iida11} S.~Iida, M.~Yukawa, H.~Yonezawa, N.~Yamamoto, and A.~Furusawa, {\it Experimental Demonstration of Coherent Feedback Control on Optical Field Squeezing}, IEEE Trans. Auto. Control. {\bf 57}, 2045 (2012).

\bibitem{Tsan10} M.~Tsang and C.~M.~Caves, {\it Coherent Quantum-Noise Cancellation for Optomechanical Sensors}, Phys. Rev. Lett. {\bf 105}, 123601 (2010).
\bibitem{Bott12} T.~Botter, D.~W.~C.~Brooks, N.~Brahms, S.~Schreppler, and D.~M.~Stamper-Kurn, {\it Linear Amplifier Model for Optomechanical Systems}, Phys. Rev. A {\bf 85}, 013812 (2012). 

\bibitem{Wise94} H.~M.~Wiseman and G.~J.~Milburn, {\it All-Optical Versus Electro-Optical Quantum-Limited Feedback}, Phys. Rev. A {\bf 49}, 4110 (1994).
\bibitem{Kerc10} J.~Kerckhoff, H.~I.~Nurdin, D.~S.~Pavlichin, and H.~Mabuchi, {\it Designing Quantum Memories with Embedded Control: Photonic Circuits for Autonomous Quantum Error Correction}, Phys. Rev. Lett.~{\bf 105}, 040502 (2010).

\bibitem{Brow07} K.~R.~Brown, J.~Britton, R.~J.~Epstein, J.~Chiaverini, D.~Leibfried, and D.~J.~Wineland, {\it Passive Cooling of a Micromechanical Oscillator with a Resonant Electric Circuit}, Phys. Rev. Lett {\bf 99}, 137205 (2007).

\bibitem{Klec06} D.~Kleckner and D.~Bouwmeester, {\it Sub-kelvin Optical Cooling of a Micromechanical
Resonator}, Nature (London) {\bf 444}, 75 (2006).
\bibitem{Gava12} E.~Gavartin, P.~Verlot, T.~J.~Kippenberg, {\it A Hybrid On-Chip Optomechanical Transducer for Ultrasensitive Force Measurements}, Nature Nanotechnol. {\bf 7}, 509 (2012).

\bibitem{Yin12} Y.~Yin {\it et al.}, {\it Catch and Release of Microwave Photon States}, Phys. Rev. Lett. {\bf 110}, 107001 (2013).
\bibitem{Reed10} M.~D.~Reed, B.~.R.~Johnson, A.~A.~Houck, L.~DiCarlo, J.~M.~Chow, D.~I.~Schuster, L.~Frunzio, R.~J.~Schoelkopf, {\it Fast Reset and Suppressing Spontaneous Emission of a Superconducting Qubit}, Appl. Phys. Lett. {\bf 96}, 203110 (2010).

\bibitem{Lloy00} S.~Lloyd, {Coherent Quantum Feedback}, Phys. Rev. A {\bf 62}, 022108 (2000).

\bibitem{Reed12} M.~D.~Reed, L.~DiCarlo, S.~E.~Nigg, L.~Sun, L.~Frunzio, S.~M.~Girvin, and R.~J.~Schoelkopf, {\it Realization of Three-Qubit Quantum Error Correction with Superconducting Circuits}, Nature (London)~{\bf 482}, 382 (2012).
\bibitem{Legh12} Z.~Leghtas, G.~Kirchmair, B.~Vlastakis, R.~Schoelkopf, M.~Devoret, and M.~Mirrahimi, {\it Hardware-Efficient Autonomous Quantum Error Correction}, arXiv:1207.0679.

\bibitem{Brag70} V.~B.~Braginskii, A.~B.~Manukin, and M.~Y.~Tikhonov, {\it Investigation of Dissipative Ponderomotive Effects of Electromagnetic Radiation }, Sov. Phys. JETP {\bf 31}, 829 (1970).
\bibitem{Giga06} S.~Gigan, H.~R.~B\"{o}hm, M.~Paternostro, F.~Blaser, G.~Langer, J.~B.~Hertzberg, K.~C.~Schwab, D.~B\"{a}uerle, M.~Aspelmeyer, and A.~Zeilinger, {\it Self-Cooling of a Micromirror by Radiation Pressure}, Nature (London) {\bf 444}, 67 (2006).
\bibitem{Arci06} O.~Arcizet, P.-F.~Cohadon, T.~Briant, M.~Pinard, and A.~Heidmann, {\it Radiation-Pressure Cooling and Optomechanical Instability of a Micromirror}, Nature (London) {\bf 444}, 71 (2006).

\bibitem{Died89} F.~Diedrich, J.~C.~Bergquist, W.~M.~Itano, and D.~J.~Wineland, {\it Laser Cooling to the Zero-Point Energy of Motion}, Phys. Rev. Lett. {\bf 62}, 403 (1989).

\bibitem{H&P} R.~L.~Hudson and K.~R.~Parthasarathy, {\it Quantum Ito's Formula and Stochastic Evolutions}, Commun. Math. Phys. {\bf 93}, 301 (1984).
\bibitem{C&G} C.~W.~Gardiner and M.~J.~Collett, {\it Input and Output in Damped Quantum Systems: Quantum Stochastic Differential Equations and the Master Equation}, Phys. Rev. A {\bf 31}, 3761 (1985).
\bibitem{QN} C.~W.~Gardiner and P.~Zoller, {\it Quantum Noise} (Springer-Verlag, Berlin, 2004).

\bibitem{Cler10} A.~A.~Clerk, M.~H.~Devoret, S.~M.~Girvin, F.~Marquardt, and R.~J.~Schoelkopf, {\it Introduction to Quantum Noise, Measurement, and Amplification}, Rev. Mod. Phys. {\bf 82}, 1155 (2010).

\bibitem{Teza11} N.~Tezak, A.~Niederberger, D.~S.~Pavlichin, G.~Sarma, and H.~Mabuchi, {\it Specification of Photonic Circuits Using Quantum Hardware Description Language}, Phil. Trans. R. Soc. A {\bf 370}, 5270 (2012).

\bibitem{Cira97} J.~I.~Cirac, P.~Zoller, H.~J.~Kimble, and H.~Mabuchi, {\it Quantum State Transfer and Entanglement Distribution among Distant Nodes in a Quantum Network}, Phys. Rev. Lett., {\bf 78} 3221 (1997).

\end{thebibliography}
\end{document}